\documentclass[%
 reprint,
 amsmath,amssymb,
 aps,
]{revtex4-2}
\usepackage{amsfonts}
\usepackage{amssymb}
\usepackage{amsthm}
\usepackage{array}
\usepackage{amsmath,mathrsfs}
\usepackage{verbatim} 
\usepackage[colorlinks=true,citecolor=blue,urlcolor=blue]{hyperref}
\usepackage{graphicx}
\usepackage{float}
\usepackage{subcaption}
\usepackage{enumerate}
\usepackage{chngcntr}
\usepackage[usenames, dvipsnames]{color}
\usepackage{graphics}
\usepackage{amsmath}

\DeclareMathOperator{\csch}{csch}

\usepackage{hyperref}
\usepackage{color}
\usepackage{bbold}
\usepackage{epstopdf}
\usepackage{mathtools}
\usepackage{enumerate}
\usepackage{subcaption}
\usepackage{physics}
\usepackage{braket}
\usepackage{amsfonts}
\usepackage{dsfont}
\usepackage{lipsum}
\usepackage[utf8]{inputenc}
\usepackage[T1]{fontenc}
\usepackage{bbold}
\usepackage{times}
\usepackage{color}
 \usepackage{hyperref} 
 \usepackage{xstring}
\hypersetup{colorlinks=true,
urlcolor=blue,
linkcolor=blue,
linktoc=page,
citecolor=blue}
\usepackage[export]{adjustbox}
\usepackage[font=footnotesize,justification=centering]{caption}
\usepackage{setspace}

\newtheorem*{proposition*}{Proposition}

\newtheorem{theorem}{Theorem}
\newtheorem*{theorem*}{Theorem}
\newtheorem{corollary}{Corollary}[theorem]
\newtheorem*{corollary*}{Corollary}

\theoremstyle{definition}

\usepackage[title,titletoc]{appendix}

\begin{document}
\title{Restoring Heisenberg-Limited Precision in Non-Markovian Open Quantum Systems via Dynamical Decoupling}
\author{Bakmou Lahcen}
\email{baqmou@gmail.com}
\affiliation{ Institute of Fundamental and Frontier Sciences,
University of Electronic Science and Technology of China, Chengdu 610051, China}
\affiliation{Key Laboratory of Quantum Physics and Photonic Quantum Information, Ministry of Education, University of Electronic Science and Technology of China, Chengdu 611731, China}

\author{Ke Zeng}
\affiliation{ Institute of Fundamental and Frontier Sciences,
University of Electronic Science and Technology of China, Chengdu 610051, China}
\affiliation{Key Laboratory of Quantum Physics and Photonic Quantum Information, Ministry of Education, University of Electronic Science and Technology of China, Chengdu 611731, China}

\author{Yu Jiang}
\affiliation{ Institute of Fundamental and Frontier Sciences,
University of Electronic Science and Technology of China, Chengdu 610051, China}
\affiliation{Key Laboratory of Quantum Physics and Photonic Quantum Information, Ministry of Education, University of Electronic Science and Technology of China, Chengdu 611731, China}

\author{Kok Chuan Tan}
\email{bbtankc@gmail.com}
\affiliation{ Institute of Fundamental and Frontier Sciences,
University of Electronic Science and Technology of China, Chengdu 610051, China}
\affiliation{Key Laboratory of Quantum Physics and Photonic Quantum Information, Ministry of Education, University of Electronic Science and Technology of China, Chengdu 611731, China}
\begin{abstract}

Non-classical resources enable measurements to achieve a precision that exceeds the limits predicted by the central limit theorem. However, environmental noise arising from system-environment interactions severely limits the performance of such resources through decoherence. While significant progress has been made in mitigating Markovian noise, the extent to which non-Markovian noise can be mitigated remains poorly understood. We demonstrate that Heisenberg Scaling, the ultimate quantum limit on measurement precision, can be recovered in quantum metrology under non-Markovian noise by leveraging carefully designed Dynamical Decoupling Techniques. Importantly, our approach does not rely on assumptions of Markovian dynamics. By imposing appropriate conditions on the control Hamiltonian, we show that HS can be achieved irrespective of whether the noise is Markovian or non-Markovian. We also prove necessary and sufficient conditions for the existence of such control Hamiltonians. As an illustrative example, we apply our framework to the damped Jaynes-Cummings model, successfully mitigating memory effects and maintaining measurement precision in complex, non-Markovian environments. These findings highlight the power of quantum control to overcome decoherence challenges and enhance metrological performance in realistic, noisy quantum systems.

\end{abstract}
\maketitle
\section{Introduction}
Quantum metrology applies quantum mechanics to sensor technology, achieving precision and sensitivity that surpass classical methods.  It has impactful applications in gravitational wave detection \cite{acernese2019increasing, tse2019quantum}, quantum clocks \cite{kruse2016improvement, nichol2022elementary}, quantum imaging \cite{dowling2015quantum}, and even emerging fields like quantum biology \cite{taylor2016quantum}. By leveraging unique quantum phenomena—such as entanglement \cite{riedel2010atom}, quantum squeezing \cite{lawrie2019quantum}, and coherence \cite{streltsov2017colloquium}—quantum metrology can enable higher precision measurements. A typical quantum metrological protocol involves three key steps: (i) Preparation of the probe, where some quantum state is prepared, (ii) Parameter-Encoding, where the probe undergoes an evolution characterized by a parameter of interest and (iii) Measurement, where the final state is measured to estimate the parameter. The precision of parameter estimation depends crucially on the probe state. If particles in the probe system are classically correlated, the uncertainty follows the Shot-Noise Limit (SNL): ${{\delta}^2} \omega \propto 1 / (N t)$, where ${{\delta}^2} \omega$ is the uncertainty in the parameter estimate, $N$ is the number of probes and $t$ is the total evolution time \cite{giovannetti2004quantum, giovannetti2006quantum, dorner2009optimal}. In contrast, non-classical resources like entanglement or coherence can achieve the Heisenberg Scaling (HS): ${{\delta}^2} \omega \propto / (N t)^2$ \cite{sanders1995optimal, zwierz2010general, demkowicz2012elusive}. This represents the ultimate precision limit attainable with quantum resources.

Achieving HS typically requires coherent state evolution, which preserves the quantum nature of the probe during the encoding process \cite{horodecki2009quantum, uhlmann1976transition}. However, realizing HS in practical systems is challenging due to decoherence. Quantum states, particularly entangled and squeezed states, are highly sensitive to noise and interactions with the environment, which leads to non-unitary evolution \cite{schleier2010states, aolita2008scaling, maccone2011beauty, huelga1997improvement}. In such open quantum systems, the precision is typically reduced to the SNL \cite{haase2016precision, smirne2016ultimate}. To combat this, various strategies have been proposed, such as quantum error correction \cite{dur2014improved, zhou2018achieving, demkowicz2017adaptive, PhysRevA.100.022312}, environmental monitoring \cite{albarelli2018restoring}, and quantum control techniques \cite{sekatski2017quantum}.

While these methods have shown promise, they often rely on the assumption of Markovian noise, where the environment does not retain memory of its interactions with the system. Such scenarios are modeled using Completely Positive (CP) maps expressed via Kraus operators \cite{kraus1983lecture, breuer2016colloquium}. CP maps can be understood as quantum operations that can be performed by acting on a joint system and environment states that are initially uncorrelated. However, in realistic systems—such as solid-state quantum platforms—non-Markovian noise can arise due to strong system-environment interactions, where the environment feeds back information to the system \cite{breuer2002theory, rivas2014quantum, breuer2009measure}. In these cases, standard CP maps may fail to describe the system dynamics accurately because the experimentalist operates may only operate on the system of interest but the total system plus environment state is initially correlated \cite{rivas2020strong, wolf2008assessing, hall2014canonical}. Non-CP maps are therefore needed to model memory effects, which are non-Markovian in nature, and often arise in solid-state quantum systems with strong interactions between the system and its environment \cite{tittel2010photon, hedges2010efficient, guo2021experimental}. This breakdown in complete positivity indicates that the system dynamics in non-Markovian processes do not conform to standard quantum channels \cite{king2001minimal, holevo1998quantum}.

Since non-Markovian noise arises due to quantum memory effects, the environment can no longer be regarded as merely a passive reservoir \cite{vasile2009continuous, liu2011experimental}. One way to model such noise is to treat it as a semi-classical stochastic process \cite{gardiner2004quantum, palma1996quantum, gardiner1985input, koch2008non}. The system's response to this noise is characterized by its noise spectrum, which describes how different frequency components affect the system \cite{breuer2002theory, clerk2010introduction, schoelkopf2008wiring}. Several types of noise spectra have been identified to describe different environmental influences on quantum systems, such as the Lorentzian noise spectrum \cite{ yu2004finite, zhang2012general}, which is typically associated with environments where a single characteristic relaxation timescale dominates the noise, and the Ohmic noise spectrum, which describes environments composed of many modes that interact with the system across a broad frequency range \cite{leggett1987dynamics, caldeira1983quantum, nagy2015nonequilibrium, prokof2000theory}.

To mitigate non-Markovian noise, advanced techniques are required. A particularly powerful method is Dynamical Decoupling Techniques (DDT) \cite{viola1999dynamical, khodjasteh2005fault, yang2011preserving, de2010universal}, which have emerged as a powerful tool for mitigating Markovian noise in quantum computing \cite{ng2011combining, piltz2013protecting, west2010high, zhang2004concatenating, matityahu2019dynamical} and quantum metrology \cite{tan2013enhancement, tan2014dephasing, sekatski2016dynamical, pham2012enhanced}. For instance, memoryless interactions with the environment can be effectively mitigated using DDT sequences like the well-known Carr-Purcell-Meiboom-Gill (CPMG) sequence \cite{carr1954effects, cywinski2008enhance, de2010universal, souza2012robust}. These sequences averages out the influence of random environmental noise over time and preserves coherent evolution. In general, DDT employs carefully designed control sequences that effectively decouple the system from its environment, preserving coherence even in noisy conditions. Unlike other methods, DDT does not require weak coupling or Markovian assumptions, making it effective for complex non-Markovian dynamics  \cite{PhysRevLett.98.100504, Biercuk_2011, PhysRevApplied.5.014007, PhysRevX.8.021059}, where memory effects and strong system-environment interactions significantly affect system behavior. Along these lines, our primary motivation is to investigate DDT as a tool in quantum metrology, where the goal is to recover HS in the presence of non-Markovian noise.

In this paper, we demonstrate that HS can be recovered in open quantum systems experiencing non-Markovian noise by employing time-dependent control Hamiltonians. Our key contributions are: (i) we derive necessary and sufficient conditions for control Hamiltonians that suppress environmental noise—Markovian or non-Markovian- while preserving a nontrivial parameter encoding, (ii) we show that, under such control Hamiltonians, the Quantum Fisher Information (QFI) scales quadratically with time ($\propto t^2$), thus achieving HS even in the presence of non-Markovian noise and, (iii) using the damped Jaynes-Cummings model with detuning frequency, we validate our findings by showing that HS is recovered using DDT. These results underscore the robustness of DDT in mitigating decoherence in quantum metrology, providing a practical framework for overcoming noise-induced challenges in quantum sensing technologies. 

The paper is organized as follows. In Sec. \ref{Sec. 2}, we outline the general framework of quantum sensing in ideal closed quantum systems. In Sec. \ref{Sec. 3}, we present the main contributions of this work: a detailed demonstration of how HS can be recovered in general open quantum systems beyond Completely Positive Trace-Preserving (CPTP) dynamical maps using DDT. This includes providing necessary and sufficient conditions for the existence of the required control Hamiltonians. In Sec. \ref{Sec. 4}, we demonstrate our findings by showing how to recover HS using DDT in the damped Jaynes-Cummings model with detuning frequency. In the final section, we summarize our results, offering concluding remarks and discussions.

\section{Preliminaries}\label{Sec. 2}

\subsection{Quantum Metrology in Closed Quantum Systems}
In metrological tasks, the goal is to estimate physical quantities such as frequency, magnetic field, or temperature, which are represented as unknown parameters. A quantum system sensitive to the parameter of interest is prepared in a specific initial state and allowed to evolve under a dynamical interaction governed by the system's Hamiltonian. The Hamiltonian encodes all the information about the parameter to be estimated. After the interaction, measurements are performed on the system's final state, and the outcomes provide the data required to estimate the parameter.

The ultimate precision for estimating a parameter $\omega$ is constrained by the Cramér-Rao bound \cite{Rao1992}: 
\begin{equation}
{\delta ^2}\hat \omega  \ge \frac{1}{{\nu {F_{\omega}}}},
\end{equation}
where $\nu$ is the number of independent measurements, and $F_{\omega}$ is the classical Fisher information, defined as:
\begin{equation}
    {F_\omega } = \int {{p_\omega }\left( x \right)} {\left( {{\partial _\omega }\log  {{p_\omega }\left( x \right)}} \right)^2}dx,
\end{equation}
where $p_{\omega}\left(x\right)$ represents the probability distribution of measurement outcomes $x$ when the parameter has value $\omega$. In the asymptotic limit, the Cramér-Rao bound can be achieved using a maximum likelihood estimation strategy \cite{Pan2002}. In quantum estimation theory, the system's initial state $\ket{\psi \left(0\right)}$, evolves to $\ket{\psi_{\omega} \left(t\right)}$ under dynamics characterized by an unknown parameter $\omega$. The precision limit in quantum measurements is determined by QFI, which provides an upper bound on the classical Fisher information over all possible measurements (POVMs). For any given pure state $\ket{\psi_{\omega} \left(t\right)}$, The QFI is expressed as \cite{braunstein1994statistical, paris2009quantum}:
\begin{equation}
    F_\omega ^{\left( Q \right)} = 4\left( {\left\langle {{\partial _\omega }{\psi _\omega }\left( t \right)} \right|\left. {{\partial _\omega }{\psi _\omega }\left( t \right)} \right\rangle  - {{\left\| {\left\langle {{\psi _\omega }\left( t \right)} \right|\left. {{\partial _\omega }{\psi _\omega }\left( t \right)} \right\rangle } \right\|}^2}} \right),
\end{equation} where $\ket{\partial_\omega\psi_\omega(t)}$ is the derivative of the quantum state with respect to the parameter $\omega$.
In closed quantum systems, the parameter $\omega$ is encoded within the system's evolution governed by the time-independent Hamiltonian $H_\omega$. The system's state at time $t$ is given by:
\begin{equation}
    \ket{\psi_{\omega} \left(t\right)}= U_{\omega}\left(t\right) \ket{\psi_{\omega} \left(0\right)}, \quad U_{\omega}\left(t\right)=\exp{\left(-i t H_{\omega}\right)},
\end{equation}
where $U_{\omega}\left(t\right)$ is the unitary time-evolution operator, and we have assumed $\hbar=1$. The unitary evolution evolves the initial state coherently, ensuring that no information about $\omega$ is lost to the environment. The QFI for the parameter $\omega$ is determined by the variance of the generator: $F_\omega^{(Q)}=4\operatorname{Var}\left[h_\omega\left(t\right)\right]_{\ket{\psi_\omega\left(t\right)}}$ where $h_\omega\left(t\right)=i \left(\partial_\omega U_{\omega}\left(t\right)\right) U_{\omega}^{\dagger}\left(t\right)$ is the generator of parameter translation with respect to $\omega$. When the Hamiltonian is a linear function of the parameter,  $H_\omega=\omega G$, the QFI simplifies to: $ F_\omega ^{\left( Q \right)}=4 t^2 \operatorname{Var}\left[G\right]_{\ket{\psi\left(0\right)}}$. The QFI achieves its maximum value when the initial state is prepared as a superposition of the eigenstates of $G$ \cite{braunstein1994statistical}: $\ket{\psi\left(0\right)}=\left(\ket{\mu_{max}}+\ket{\mu_{min}}\right)/\sqrt{2}$, where $\mu_{max}$ and $\mu_{min}$ are the maximal and minimal eigenvalues of $G$ with corresponding eigenstates $\ket{\mu_{max}}$ and $\ket{\mu_{min}}$. In this case, the maximum QFI is: $ F_\omega ^{\left( Q \right)}=t^2 \left(\mu_{max}-\mu_{min}\right)^2$. When these conditions are satisfied, we see that the system achieves HS with respect to time, with the QFI scaling as $t^2$.

\subsection{Quantum Metrology in Open Quantum Systems}

The steps of a metrological protocol in open quantum systems are similar to those in closed systems. However, a key distinction lies in the interaction between the system of interest and its environment, which introduces complexities. As a result, the system’s dynamical evolution is no longer described by a unitary operator. Instead, the total Hilbert space expands to include both the system and the environment: $\mathcal{H}_{tot}=\mathcal{H}_{S} \otimes  \mathcal{H}_{E}$, where $\mathcal{H}_{S}$ and $\mathcal{H}_{E}$ represent the Hilbert spaces of the system and environment, respectively.  In this framework, the joint system is treated as a closed system governed by a total Hamiltonian:
\begin{equation}\label{Eq. 5}
    H_{tot}=H_S\left(\omega\right) \otimes \mathbb{1}_E+\mathbb{1}_S\otimes H_E+H_{SE},
\end{equation}
where $H_S(\omega)$ is the system’s Hamiltonian, encoding the parameter $\omega$ to be estimated. The environment is described by $H_E$, often referred to as the reservoir Hamiltonian, while $H_{SE}$ captures the interaction between the system and environment, responsible for decoherence. The identity operators $\mathbb{1}_S$ and $\mathbb{1}_E$ act on the system and environment Hilbert spaces, respectively. 

To account for interactions with the environment, the system’s time evolution is best described using density matrices. The joint system evolves unitarily under the total Hamiltonian $H_{\text{tot}}$, with the time-evolution operator: $ U_{tot}\left(\omega,t\right) = \exp[-i t H_{\text{tot}}]$. The von Neumann equation governs this evolution:
\begin{equation}
    \frac{{d{\rho _{SE}}\left( t \right)}}{{dt}} =  - i\left[ {{H_{tot}},{\rho _{SE}}\left( t \right)} \right],
\end{equation}
yielding the formal solution: $\rho_{SE}\left(t\right) =U_{tot}\left(\omega,t\right)\rho_{SE}\left(0\right)U_{tot}^{\dagger}\left(\omega,t\right)$.  The reduced density matrix of the system is obtained by tracing out the environmental degrees of freedom:
\begin{equation}
    \rho_S\left(\omega, t\right) = \operatorname{Tr}_E\left[U_{tot}\left(\omega,t\right)\rho_{SE}\left(0\right)U_{tot}^{\dagger}\left(\omega,t\right)\right].
\end{equation}
This reduced density matrix captures the system’s dynamics under environmental influence, and is generally a mixed state due to entanglement with the environment. The process of entanglement with the environment leads to information about the parameter $\omega$ being distributed across the system-environment composite. Consequently, the reduced state $\rho_S\left(\omega, t\right)$ cannot retain all the information, resulting in a degraded QFI. The QFI for $\omega$ in the reduced density matrix is: $F_{\omega}^{(Q)} = \operatorname{Tr}[\rho_S\left(\omega, t\right) L^2\left(\omega, t\right)]$,  where $ L\left(\omega, t\right)$ is the Symmetric Logarithmic Derivative (SLD) operator, defined by $2 \partial_\omega \rho_S\left(\omega, t\right) = L\left(\omega, t\right) \rho_S\left(\omega, t\right) + \rho_S\left(\omega, t\right) L\left(\omega, t\right)$. As the system’s correlation with the environment increases, the precision of estimating $\omega$ decreases, since more information about $\omega$ is lost to the environment. However, because the joint system evolves unitarily, the total information is encoded coherently in the global pure state. Thus, the QFI of $\omega$ in the reduced state is always bounded by the QFI of $\omega$ in the global pure state:
\begin{equation}
    F_{\omega}^{(Q)}\left(\rho_S\left(\omega,t\right)\right) \leqslant F_{\omega}^{(Q)}\left(\ket{\psi_{\omega}\left(t\right)}_{SE}\right).
\end{equation}
This inequality can be expressed as:
\begin{equation}\label{Ineq. 9}
   F_{\omega}^{(Q)}\left(\rho_S\left(\omega,t\right)\right) \leqslant  4 \operatorname{Var}\left[h_\omega\left(0 \rightarrow t\right)\right]_{\ket{\psi\left(0\right)}_{SE}},
\end{equation}
where $h_\omega\left(0 \rightarrow t\right)$ is the generator associated with the global unitary evolution, defined as: $h_\omega\left(0 \rightarrow t\right)=i U_{tot}^{\dagger}\left(\omega,t\right) \partial_{\omega}U_{tot}\left(\omega,t\right)$ \cite{Pang2017}. The degradation of QFI in $\rho_S(\omega, t)$ stems from decoherence caused by the exchange of information between the system and the environment. To recover HS, effective quantum control must: (i) suppress detrimental system-environment interactions, and (ii) preserve the system’s ability to encode $\omega$. In the following sections, we present the theoretical framework and control strategies needed to achieve this, enabling the recovery of Heisenberg scaling (HS) of precision.

\section{Results}\label{Sec. 3}

We now address the central focus of this work: exploring quantum metrology within the framework of open quantum systems. Unlike idealized closed systems, real-world quantum systems interact with their surroundings, resulting in non-unitary evolution. Such interactions lead to decoherence, where the quantum properties critical for high-precision metrology degrade over time.

Our primary objective is to investigate the use of DDT methods for preserving Heisenberg scaling, even under the challenges posed by non-unitary dynamics, including non-Markovian dynamics. This requires developing robust strategies to mitigate the detrimental effects of the environment, ensuring that the system retains its quantum features sufficiently to achieve precision beyond SNL.

\subsection{Restoring HS in Open Quantum Systems via DDT }

To recover the HS of precision in noisy quantum systems, we will employ DDT to address the challenge of controlling the generator $h_\omega\left(0 \rightarrow t\right)$, thereby mitigating the detrimental effects of system-environment interactions. A key strength of DDT is its applicability beyond standard CP maps, making it effective for general dynamical maps, including those involving complex non-Markovian noise. 

The central idea is to construct a time-dependent control Hamiltonian $H_C(t)$ that acts only on the system, but is able to counteract the interaction with the environment. Under this approach, the total Hamiltonian becomes:
\begin{equation}\label{Eq. 10}
 \small   H_{tot}\left(t\right)= H_S\left(\omega\right)\otimes \mathbb{1}_E+ H_C\left(t\right)\otimes \mathbb{1}_E+\mathbb{1}_S\otimes H_E+H_{SE},
\end{equation}
where $H_C\left(t\right)$ acts only on the Hilbert subspace $\mathcal{H}_S$ and its evolution is governed by ${U_C}\left( t \right) = {\mathcal{T}}\exp\left( { - i\int\limits_0^t {ds{H_C}\left( s \right)} } \right)$, where $\mathcal{T}$ denotes the time-ordering operator. In the control frame, the time evolution of the global state is described by the unitary operator: ${\tilde  U_{tot}}\left( \omega, t \right) = {\mathcal{T}}\exp\left( { - i\int\limits_0^t {ds{\tilde  H_{tot}}\left( s \right)} } \right)$. Here, the generator of the parameter $\omega$ is rewritten as $\tilde h_\omega\left(0 \rightarrow t\right)=i \tilde  U_{tot}^{\dagger}\left(\omega,t\right) \partial_{\omega} \tilde U_{tot}\left(\omega,t\right)$ and the Hamiltonian in the control frame is expressed as ${\tilde H_{tot}}\left( t \right) = {U_C}\left( t \right){H_{tot}(t)}U_C^\dagger \left( t \right)$.  The total Hamiltonian $\tilde  H_{tot}\left(t\right)$ now becomes time-dependent due to $H_C(t)$. To simplify this evolution, we approximate it as a sequence of discrete, time-independent intervals by dividing the evolution into $n = t/T$ small intervals of duration $T$, where $T \rightarrow 0$. The total unitary evolution can be broken into $n$ unitaries evolutions:
\begin{eqnarray}\label{Eq. 11}
  \begin{aligned}
        {\tilde  U_{tot}}\left(\omega, t \right)= & {\tilde  U_{tot}}\left(\omega,t-T \to t\right)  
     \\ & \cdots {\tilde  U_{tot}}\left(\omega,2T \to T\right) {\tilde  U_{tot}}\left(\omega,0 \to T\right). 
    \end{aligned}
\end{eqnarray}
When the control time interval $T$ is sufficiently short, each interval is approximately time independent \cite{Pang2017}. The unitary evolution operator over each interval can be expressed as:
\begin{equation}
    \tilde U_{tot}\left(\omega, kT \rightarrow (k+1)T\right) \approx e^{-i T \tilde H_{tot}\left(kT\right)},
\end{equation}
 where $\tilde H_{tot}(kT)$ is the Hamiltonian at the beginning of each interval labeled by $k$. The total unitary evolution over $n$ intervals is approximated as:
 \begin{eqnarray}\label{Eq. 13}
    \begin{aligned}
         \tilde U_{tot}\left(\omega,t \right) \approx   e^{ - iT\sum\nolimits_{k = 0}^{n - 1} {{\tilde H_{tot}}\left( {kT} \right)}}.
    \end{aligned}
\end{eqnarray}
 This approximation becomes arbitrarily accurate in the limit of $T \rightarrow 0 $, or equivalently, $n \rightarrow \infty $. Substituting the effective Hamiltonian components, this becomes:
\begin{equation} \label{Eq. 14}
    {\tilde U_{tot}}\left( {\omega ,0 \to nT} \right) \approx \quad {e^{ - inT\left( {H_S^{\text{eff}} \otimes {\mathbb{1}_E} + {\mathbb{1}_S} \otimes H_E + H_{SE}^{\text{eff}}} \right)}},
\end{equation}
where $H_S^{\text{eff}}\left(\omega\right) = \frac{1}{n}\sum\nolimits_{k = 0}^{n - 1} {U_C\left( {kT} \right){H_S}\left( \omega  \right)U_C{{\left( {kT} \right)}^\dagger}}$,   and $H_{SE}^{\text{eff}} = \frac{1}{n}\sum\nolimits_{k = 0}^{n - 1} {\left( {U_C\left( {kT} \right) \otimes {\mathbb{1}_E}} \right){H_{SE}}\left( {U_C{{\left( {kT} \right)}^\dagger } \otimes {\mathbb{1}_E}} \right)}$. 

To achieve decoupling, the control unitaries $U_C(kT)$ for each time interval labeled by $k$ are designed to satisfy the dynamical decoupling condition:
\begin{equation}\label{Eq. 15}
   \frac{1}{n}\sum\limits_{k=0}^{n-1} {\left( {{U_C}(kT) \otimes \mathbb{1}_E} \right)} {H_{SE}}\left( {U_C^\dagger (kT) \otimes \mathbb{1}_E} \right) =  c  \mathbb{1}_S \otimes {J_E},
\end{equation}
where $c$ is some constant, and $J_E$ is a Hermitian operator acting on the environment Hilbert space. To perform parameter estimation, we further require that the effective system Hamiltonian $H_S^{\text{eff}}\left(\omega\right)$ generates nontrivial evolution, which can be expressed as:
\begin{equation} \label{Eq. 15-2}
    \frac{1}{n}\sum\nolimits_{k = 0}^{n - 1} {U_C\left( {kT} \right){H_S}\left( \omega  \right)U_C{{\left( {kT} \right)}^\dagger}} \not\propto \openone_S.
\end{equation}

Eqs.~\ref{Eq. 15} and \ref{Eq. 15-2} ensure that the system is effectively decoupled from the environment, suppressing noise while preserving the system’s ability to encode the parameter $\omega$. The objective is to find a set of quantum control operations $\{ U_C(kT) \}_{k=0}^{n-1}$ that satisfies both these conditions. The following theorem proves necessary and sufficient conditions for the existence of such quantum control.

\begin{theorem} [Necessary and sufficient conditions for dynamical decoupled quantum sensing] \label{th::iff}
    For any system-environment interaction $H_S \otimes \mathbb{1}_E + \mathbb{1}_S \otimes H_E + H_{SE}$, there exists a collection of unitary operators $\{ {U_C(kT)} \}_{k=0}^{n-1}$ that satisfies 
    \begin{equation*}
        H^{\text{eff}}_S = \frac{1}{n}\sum_{k=0}^{n-1} U_C(kT) \, H_S \, U_C(kT)^\dagger \not\propto \mathbb{1}_S
    \end{equation*} 
    and 
    \begin{align*}
       H_{SE}^{\text{eff}} &= \frac{1}{n}\sum_{k=0}^{n-1}\left( U_C(kT)\otimes \mathbb{1}_E \right) H_{SE} \left(U_C(kT)^\dagger \otimes \mathbb{1}_E\right) \\&= c \mathbb{1}_S \otimes J_E
    \end{align*}
    for some constant $c$ and Hermitian operator $J_E$, if and only if there exists mixed unitary channel $\Phi$ such that $\Phi(H_S)$ is nontrivial,
  \begin{equation*}
      \operatorname{Tr} \left\{\left[ \Phi(H_S) - \operatorname{Tr}(H_S) \frac{\mathbb{1}_S}{d} \right] \Phi \left[ \prescript{}{E}{\braket{\psi|H_{SE}|\psi}_E} \right] \right\}= 0
  \end{equation*}  
    and 
    \begin{equation*}
        \braket{i | \Phi \left[ \prescript{}{E}{\braket{\psi|H_{SE}|\psi}_E} \right]| i} = \mathrm{constant},
    \end{equation*}
    for every eigenvector $\ket{i}$ of $\Phi(H_S)$ and state vector $\ket{\psi}_E$.
\end{theorem}

A detailed proof of this Theorem is provided in Appendix~\ref{App. A}. By satisfying the specified conditions, the interaction term $H_{SE}$ is effectively averaged out, decoupling the system's dynamics from the environment. This ensures that no nontrivial system-dependent terms remain in the interaction Hamiltonian, allowing the system to evolve independently of the environment. As a special case, when $c=0$, the summation-based condition retrieves the integral condition established in \cite{facchi2005control}. The integral condition requires periodicity in the control unitaries to average the interaction Hamiltonian to zero over a complete cycle. In contrast, our summation-based condition relaxes this requirement by averaging the interaction over discrete, potentially non-periodic control intervals, thus covering more general physical scenarios. 

Since the identity channel $\Phi(\cdot) = (\cdot)$ , which leaves the input operator unchanged, is a special case of a mixed unitary operator, the following corollary provides a simpler sufficient condition expressed in terms of the initial input Hamiltonian:

\begin{corollary}[Sufficient conditions]\label{cor::if}
    Dynamically decoupled quantum sensing in the sense of Theorem~\ref{th::iff} is possible if $H_S$ is nontrivial,  
    \begin{equation*}
        \operatorname{Tr} \left\{\left[ H_S - \operatorname{Tr}(H_S) \frac{\mathbb{1}_S}{d} \right]   \prescript{}{E}{\braket{\psi|H_{SE}|\psi}_E}  \right\}= 0
    \end{equation*}
    and 
    \begin{equation*}
        \braket{i | \prescript{}{E}{\braket{\psi|H_{SE}|\psi}_E} | i} = \mathrm{constant},
    \end{equation*}
    for every eigenvector $\ket{i}$ of $H_S$ and state vector $\ket{\psi}_E$.
\end{corollary} Corollary~\ref{cor::if} asserts that it is possible to dynamically decouple the system from the environment while ensuring that the effective system Hamiltonian, $H_S^{\text{eff}}$, remains non-trivial. This is achievable whenever the noise introduced by the environment, characterized by $\prescript{}{E}{\braket{\psi|H_{SE}|\psi}_E}$, is orthogonal to the signal-generating Hamiltonian, represented by the non-trivial portion of $H_S$. 

Upon satisfying this condition in Theorem~\ref{th::iff}, the total unitary evolution from $t=0$ to $t=nT$ can be rewritten as:
\begin{eqnarray}\label{Eq. 16}
    \begin{aligned}
        {\tilde  U_{tot}}\left(\omega, 0  \rightarrow n T\right) & 
        \approx {e^{ - inT\left( {{H_S^{\text{eff}}} \otimes {\mathbb{1}_E} + {\mathbb{1}_S} \otimes \left( {{H_E} + c{J_E}} \right)} \right)}}\\&
        \approx U_S^{\text{eff}}\left(\omega,t\right)\otimes U_E^{\text{eff}}\left(t\right),
    \end{aligned}
\end{eqnarray}\\
where $U_S^{\text{eff}}\left(\omega, t\right)=e^{-i t H_S^{\text{eff}}\left(\omega\right)}$, $U_E^{\text{eff}}\left(\omega, t\right)=e^{-i t H_E^{\text{eff}}}$ and $H_E^{\text{eff}}=H_E+c J_E$. This shows that the system and environment evolve independently, with $U_S^{\text{eff}}\left(\omega, t\right)$ governing the effective dynamics of the system. Using the effective dynamics, the generator of parameter $\tilde h_\omega\left(0 \rightarrow t\right)$ is expressed as:
\begin{eqnarray}
   \small{ \begin{aligned}\label{Eq. 17}
\tilde h_\omega\left(0 \rightarrow t\right)  &= i \tilde  U_{tot}^{\dagger}\left(\omega,0 \rightarrow t\right) \partial_{\omega} \tilde U_{tot}\left(\omega,0 \rightarrow t\right)\\&
\approx i U_S^{\text{eff} \hspace{0.1cm} \dagger}\left(\omega,t\right)\partial_{\omega}U_S^{\text{eff}}\left(\omega,t\right) \otimes \mathbb{1}_{E}\\&
=  h_{\omega}^{\text{eff}}\left(0 \rightarrow t\right) \otimes \mathbb{1}_{E},
    \end{aligned}}
\end{eqnarray}
where  $ h_{\omega}^{\text{eff}}\left(0 \rightarrow t\right)=i U_S^{\text{eff}  \hspace{0.1cm} \dagger}\left(\omega, t\right)\partial_{\omega}U_S^{\text{eff}}\left(\omega, t\right)$ is the effective generator after the quantum controls have been applied. Since $\tilde{h}_\omega\left(0 \rightarrow t\right)$ acts exclusively on the system state, the environment no longer interferes with the encoding of the parameter $\omega$.  

If the system Hamiltonian is linear in $\omega$, such that $H_S\left(\omega\right) = \omega G_S$, the effective Hamiltonian under dynamical decoupling is also linear, $H^{\text{eff}}_S(\omega)=\omega G_S^{\text{eff}}=\frac{\omega}{n}\sum_{k=0}^{n-1}U_C(kT) G_SU_C^{\dagger}(kT)$. In this case, one can easily calculate the effective generator and obtain:
\begin{equation}
    h^{\text{eff}}_{\omega}\left(0 \rightarrow t\right)= t\ G_S^{\text{eff}}=\frac{t}{n} \sum\nolimits_{k=0}^{n-1} {U_C(kT) G_SU_C^{\dagger}(kT)}.
\end{equation}
For a unitary evolution of the form $e^{-i \omega A }$, QFI is given by: 
\begin{equation}\label{Eq. QFIRHO}
    F_{\omega}^{(Q)} (\rho_{S}(\omega,t))= 2 \sum_{k \ne l} \frac{(\lambda_k - \lambda_l)^2}{(\lambda_k + \lambda_l)} \abs{\braket{\psi_k|A|\psi_l}}^2,
\end{equation}
where $\lambda_k$ and $\ket{\psi_k}$ are the eigenvalues and eigenvectors of $\rho_{S}(\omega,t)$ respectively. In our case, the unitary evolution of the system is described by $e^{-i \omega t G_S^{\text{eff}} }$, with $A \rightarrow t G_S^{\text{eff}} $. Substituting this into the QFI expression yields:
\begin{equation}
    F_{\omega}^{(Q)} (\rho_{S}(\omega,t)) \propto t^2.
\end{equation}

This scaling demonstrates that the QFI grows quadratically with time, which is the HS. This confirms that dynamic decoupling enables the preservation of HS even in noisy quantum systems, provided the effective Hamiltonian $G_S^{\text{eff}}$ is properly constructed according to the proper requirements (see Theorem~\ref{th::iff}). Significantly, note that the recovery of HS does not assume that the system-environment interaction is Markovian in nature.

For a specific case where the states of the system and environment are initially uncorrelated (separable), $\ket{\psi\left(0\right)}_{SE} = \ket{\varphi\left( 0 \right)}_S\otimes\ket{\chi \left( 0 \right)}_E$, the QFI can be evaluated as:
    \begin{equation}
         F_\omega ^{(Q)}\left( \rho _S\left( \omega ,t \right)\right) \approx \ 4 \ t^2 \operatorname{Var} \left[ G^{\text{eff}} \right]_{\ket{\varphi\left( 0 \right)}_S}.
    \end{equation}
Here, $ \rho _S\left( \omega,t \right)$ represents the density matrix of the pure state of the system, as it is effectively decoupled from the environment. Therefore, the expression for QFI is similar to the noiseless case, as discussed in the first subsection of Sec. \ref{Sec. 2}.

\section{Example: Frequency estimation in the presence of non-Markovian noise} \label{Sec. 4}

To demonstrate the applicability of our results, we consider the problem of frequency estimation in the presence of non-Markovian noise. Frequency measurement is a cornerstone of quantum metrology, with applications spanning diverse scientific fields, including atomic clock technology and gravitational wave detection\cite{acernese2019increasing, tse2019quantum, kruse2016improvement, nichol2022elementary}.  Among various quantum platforms, single-phase estimation has emerged as a prominent technique for achieving high-precision frequency measurements. This approach has been experimentally realized in systems such as single nuclear spins in diamond, particularly those employing nitrogen-vacancy (NV) centers \cite{waldherr2012high}.

In this section, we investigate a specific example: estimating the frequency of two-level atoms embedded within a cavity coupled to $k$-harmonic oscillators. We compare the QFI with and without the application of dynamical decoupling, highlighting the role of DDT in enhancing quantum metrology under non-Markovian noisy conditions.

 \subsection{Frequency estimation without DDT}

We consider a two-level atom (qubit) oscillating between its ground ($\ket{g}$) and excited ($\ket{e}$) states with frequency $\omega_0$, embedded within a photonic cavity coupled to a $k$-chain of quantum harmonic oscillators with frequencies $\omega_k$. The task is to estimate $\omega_0$. In the rotating wave approximation, the total Hamiltonian of the system is given by:
\begin{equation}
    H_{tot}=\frac{\omega_0}{2}\sigma_{z}+\sum\nolimits_{k} \omega_k a_k^{\dagger}a_k + \left(\sigma_{+}B +\sigma_{-}B^{\dagger}\right),
\end{equation}
where $a_k$ and $a_k^{\dagger}$ are the annihilation and creation operators for the photonic field, $\sigma_z$, $\sigma_{+}$, and $\sigma_{-}$ are the Pauli $Z$, raising, and lowering operators of the qubit, $g_k$ are coupling strengths with the $k$th mode, and $B \coloneqq \sum_k g_k a_k$.

The coupling strength $g_k$ is related to the reservoir’s spectral density $J\left(\omega\right)$, which, in the continuum frequency limit, is defined as $J(\omega) = \sum_k |g_k|^2 \delta(\omega - \omega_k)$ defines how the coupling varies across frequencies. For this model, we assume a Lorentzian spectral density with detuning:
\begin{equation}
    J(\omega) = \frac{1}{2\pi}\frac{\gamma_0 \lambda^2}{(\omega - \omega_0 - \Delta)^2 + \lambda^2},
\end{equation}
  where $\gamma_0$ is the coupling rate at resonance, $ \Delta = \omega_0 - \omega_c $ is the detuning, with $ \omega_c $ as the cavity’s center frequency, and $\lambda$ defines the spectral width and is inversely related to the reservoir correlation time $ \tau_R = 1/\lambda $. A large detuning $ \Delta $ reduces the effective coupling between the qubit and its environment, thus decreasing the reservoir's influence on the qubit's dynamics. The reservoir exhibits non-Markovian behavior when $ \lambda $ is small (longer correlation times), and Markovian behavior when $ \lambda $ is large (short correlation time). Non-Markovian effects are particularly prominent when $\gamma_0  \gg  \lambda$~\cite{lu2010quantum}.
  
    In the absence of noise, QFI for the parameter $\omega_0$ is $F = 4t^2 \operatorname{Var}[\sigma_z]_{\ket{\varphi\left(0\right)}}$. where $\ket{\varphi\left(0\right)}_S = \left( \ket{e} + \ket{g} \right)/\sqrt{2}$  is the optimal probe state. This state maximizes $\operatorname{Var}[\sigma_z]_{\ket{\varphi\left(0\right)}}$, leading to $F_{\omega_0}^{(Q)} = t^2$. This corresponds to HS, the precision limit in quantum metrology.

Now consider the presence of noise, with the qubit initially prepared in the state $\ket{\varphi\left(0\right)}_S$. As the system evolves under the total Hamiltonian, the state at time $t$ remains confined to the single-excitation subspace and takes the form: 
\begin{equation}
       \ket{\psi(t)}_{SE} = C_e(t) \ket{e 0} + C_g(t) \ket{g 0} + \sum\nolimits_k {C_k(t) \ket{g 1_k}},
\end{equation}
where $C_e(t)$, $C_g(t)$, and $C_k(t)$ are time-dependent amplitudes, $\ket{e 0}$ and $\ket{g 0}$ are the excited and ground states with no photons in the cavity, and $\ket{g1_k}$ denotes the qubit in the ground state with one photon in the $k$-th cavity mode. In general, this is an entangled state so $\ket{\psi(t)}_{SE}$ cannot be expressed as a tensor product of system and environment states. These correlations lead to quantum memory effects and non-Markovian noise. The evolution of the global system state $\ket{\psi(t)}_{SE}$ is governed by the Schrödinger equation $i \partial_t\ket{\psi(t)}_{SE}=H_{tot}\ket{\psi(t)}_{SE}$.  Solving this equation yields the state at time $t$ (see  Appendix~\ref{App. C} for detailed calculations):
\begin{eqnarray}
    \begin{aligned}
         \ket{\psi\left(t\right)}_{SE}=& e^{-i\frac{\omega_0}{2} t}c_e(t)\ket{e 0}+\frac{e^{i\frac{\omega_0}{2}}}{\sqrt{2}}\ket{g 0} \\&
         +\sum\nolimits_k {e^{-i\left(\omega_k-\frac{\omega_0}{2}\right)} c_k(t)\ket{g 1_k}}.
    \end{aligned}
\end{eqnarray}
To obtain the reduced state of the qubit, we trace out the environmental degrees of freedom:
\begin{equation}
    \rho _{S}\left(\omega_0, t \right) = \left( {\begin{array}{*{20}{c}}
{{{\left| {{c_e}\left( t \right)} \right|}^2}}&{\frac{1}{{\sqrt 2 }}{e^{ - i{\omega _0}t}}{c_e}\left( t \right)}\\
{\frac{1}{{\sqrt 2 }}{e^{i{\omega _0}t}}c_e^*\left( t \right)}&{1 - {{\left| {{c_e}\left( t \right)} \right|}^2}}
\end{array}} \right).
\end{equation} \
This reduced density matrix captures the qubit's dynamics, reflecting the influence of the environment. The QFI for estimating a parameter $\omega_0$ from a density matrix $\rho _S(\omega_0 ,t)$ is computed using the formula:
\begin{equation}
F_{\omega_0} ^{(Q)} = \sum\limits_k {\frac{{{{\left( {{\partial _{\omega_0} }{\lambda _k}} \right)}^2}}}{{{\lambda _k}}}}  + 2\sum\limits_{k \ne l} {\frac{{{{\left( {{\lambda _k} - {\lambda _l}} \right)}^2}}}{{{\lambda _k} + {\lambda _l}}}} {\left| {\langle {\psi _k}|{\partial _{\omega_0} }{\psi _l}\rangle } \right|^2}.
\end{equation}
This expression represents the general form of \ref{Eq. QFIRHO}, in which the dynamic generator fully captures the information of $\omega_0$.

\begin{figure}[H]
\includegraphics[scale=0.66]{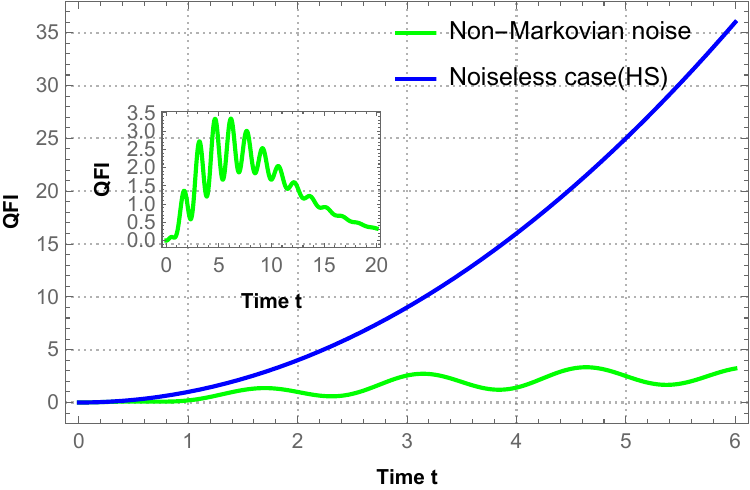}   
    \caption{Plot of QFI as a function of rescaled time $t$ for frequency estimation in a two-level system embedded in a $k$-oscillator harmonic cavity, modeled by a Lorentzian spectral density with detuning. The parameters used in this simulation are $\lambda=0.5$, $\gamma_0=10 \lambda$ and $\Delta=3 \lambda$.}
    \label{Fig. 1}
\end{figure}

The results for QFI are presented in  Fig.~\ref{Fig. 1}, which compares the noiseless and noisy cases. The blue curve in Fig.~\ref{Fig. 1} represents the QFI for the noiseless case. In this scenario, the QFI scales as $t^2$, corresponding to HS. The green curve in Fig.~\ref{Fig. 1} shows the QFI for the noisy case. Here, the QFI exhibits oscillations as a function of time $t$, rather than monotonically increasing. These oscillations arise from the strong system-environment interactions that introduce significant memory effects, characteristic of non-Markovian dynamics. Non-Markovian noise allows the environment to retain information about the system's past interactions and feed it back. For example, photons emitted by the qubit can be temporarily stored in the cavity and later reabsorbed by the system. When information flows back into the qubit, the QFI increases, enhancing the system's sensitivity to frequency estimation. Conversely, when information flows from the qubit into the environment, the QFI decreases as the system loses coherence. This alternating information exchange leads to the observed QFI oscillations, where the system periodically gains and loses sensitivity to the estimated parameter.

\begin{figure}[H]
\includegraphics[scale=0.67]{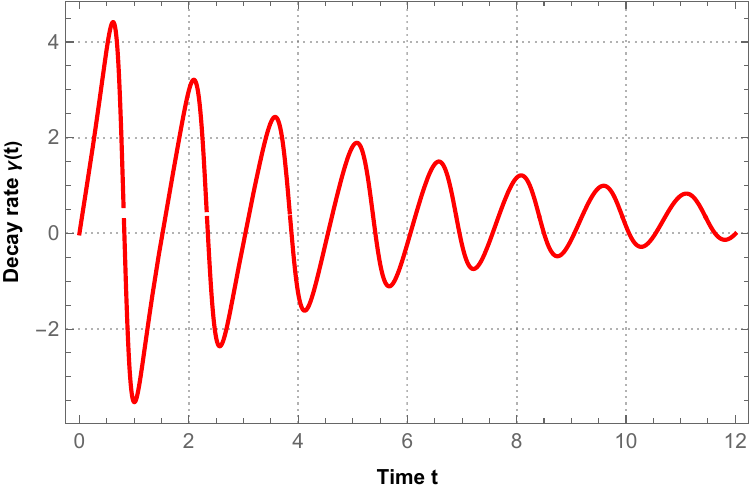}   
    \caption{Plot of the decay rate as a function of rescaled time $t$ for a two-level atom coupled to a reservoir described by a Lorentzian spectral density with frequency detuning. The simulation parameters are $\lambda=0.5$, $\gamma_0=10 \lambda$ and $\Delta=3 \lambda$.} \label{Fig. 2}
\end{figure}

We plot the decay rate as a function of time $t$ (see Fig.~\ref{Fig. 2}). The decay rate exhibits oscillations with both positive and negative values. Positive decay rates indicate coherence loss, as the system becomes more entangled with the environment. Negative decay rates reflect coherence regrowth as the system recovers information from the environment. These oscillations in the decay rate are a hallmark of non-Markovian dynamics \cite{breuer2016colloquium, breuer2002theory, rivas2014quantum}, where the system's evolution depends not only on its current state but also on its history.

\subsection{Frequency estimation with DDT} 

Now, to mitigate this type of noise, we employ DDT by introducing an additional time-dependent control Hamiltonian to the original global Hamiltonian. Specifically, we choose $H_C(t) = \frac{\pi}{2} \sum_{k=0}^{n-1} \delta(t - kT) \sigma_z$,
where the delta function $\delta(t - kT)$ acts at discrete time intervals and applies a $\pi/2$ rotation to the target system via the control unitary evolution operator $ U_C(t)=\mathcal{T}\exp{\left(-i\int_0^tdsH_C(s)\right)}$~\cite{tan2013enhancement}. The global Hamiltonian then becomes:
\begin{equation}\label{Eq. HDDT}
  \small  H_{tot}(t)=\frac{\omega_0}{2}\sigma_{z}+\sum_{k} \omega_k a_k^{\dagger}a_k + \left(\sigma_{+}B +h.c\right)+\frac{\pi}{2} \sum_{k=0}^{n-1} \delta(t - kT) \sigma_z.
\end{equation}
It can be readily verified that the control unitary evolution satisfies the DDT condition in Eq. \ref{Eq. 15}. This ensures that the interaction term is effectively canceled out over the entire sequence of control operations, thereby decoupling the system from environmental noise. 

In the control picture, the Hamiltonian in Eq. \ref{Eq. HDDT}, can be rewritten as:
\begin{equation}
\small \tilde H_{tot}(t)=\frac{\omega_0}{2}U_C(t)\sigma_{z}U_C^{\dagger}(t)+\sum_{k} \omega_k a_k^{\dagger}a_k + \left(-1\right)^n\left(\sigma_{+}B +h.c\right),
\end{equation}
where $n=t/T$ denotes the number of control operations applied up to time $t$. Using the same initial state preparation and following the same approach as in the case without DDT, one derives the differential equation for the time-dependent amplitude of the excited state:
\begin{equation}\label{Eq. Dce}
\dot c_e(t)=-\int_0^{t} f\left(t-\tau\right) \left(-1\right)^{n+\tau/T} c_e(\tau),
\end{equation}
where the correlation function takes the form:
\begin{equation}
    f(t-\tau) = \frac{1}{2} \gamma_0 \lambda \left(-1\right)^{n+\tau/T} e^{-(\lambda - i \Delta)(t - \tau)}.
\end{equation}
Differentiating Eq.~\ref{Eq. Dce} with respect to $t$ over the interval $t \in \left[(n-1)T, nT\right]$ yields the second-order equation of motion: 
\begin{equation}\label{Eq. SecondDce}
\ddot c_e(t) =  - \frac{1}{2}{\gamma _0}\lambda {c_e}(t) + \left( {i\Delta  - \lambda } \right){\dot c_e}(t).
\end{equation}
This equation is a second-order differential equation with complex coefficients and admits the general solution:
\begin{equation}\label{Eq. Soce}
{c_e}\left( t \right) = {c_e}\left( 0 \right){e^{ - \frac{\delta }{2}t}}\left( {{A_n}\cosh \left[ \frac{d \tau_n}{2} \right] + {B_n}\sinh \left[  \frac{d \tau_n}{2} \right]} \right),
\end{equation}
where $\delta=\lambda-i\Delta$ and $\tau_n=t-nT$. Explicit expressions for the coefficients $A_n$ and $B_n$ are given in Appendix~\ref{App. D}. In the resonance case, $\Delta=0$, and when $n=0$, it can be easily shown that $c_e(t)$ reduces to Eq. (10.45) in Ref. \cite{breuer2002theory}.

\begin{figure}[H]
\includegraphics[scale=0.67]{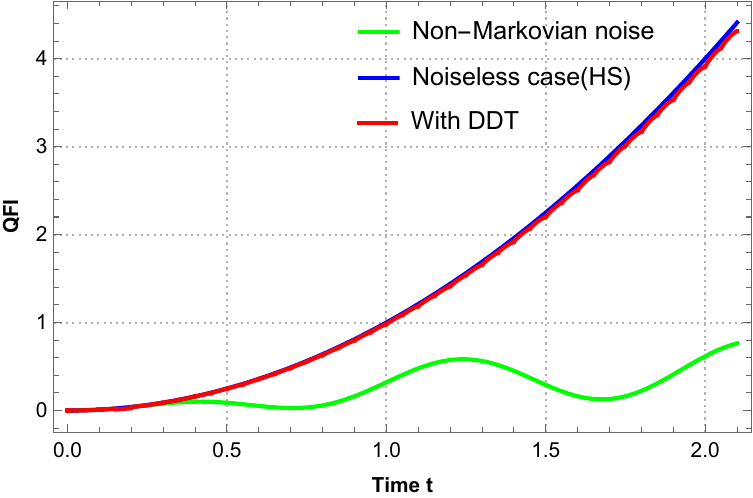}   
    \caption{QFI for frequency $\omega_0$ as a function of rescaled time $t$, for a two-level atom coupled to a reservoir with a Lorentzian spectral density. The parameters are set to $\lambda=0.5$, $\gamma_0=10 \lambda$ and $\Delta=3 \lambda$.}\label{Fig. 3}
\end{figure}

Transforming back to the Schrödinger picture and following the same procedure as in without DDT case, we obtain the results shown in Fig.~\ref{Fig. 3}, which illustrates the QFI for the frequency $\omega_0$ as a function of time $t$ under three scenarios: non-Markovian noise without DDT (green), non-Markovian noise with DDT (red), and a noiseless system (blue). The green curve shows that non-Markovian noise degrades the QFI over time through continuous interactions between the system and its environment. The oscillations in QFI reflect the dynamic exchange of information between the two-level atom and the photonic cavity, as photons are emitted and then reabsorbed. In contrast, the red curve, corresponding to sequential control pulses under DDT, exhibits a significant recovery of the QFI, closely approaching the Heisenberg scaling observed in the noiseless case (blue). This underscores DDT’s effectiveness in preserving coherence and improving estimation precision. The small remaining gap between the red and blue curves is attributed to practical limitations: a finite number of control pulses and the time each operation requires. However, increasing the number of pulses or decreasing the interval between them can further restore ideal precision, regardless of whether the noise is Markovian or non-Markovian. This highlights the critical importance of optimizing DDT for real-world applications.
\section{Conclusion and outlook}

We have demonstrated that DDT offers a robust strategy for enhancing measurement precision in quantum metrology, even under the challenging conditions imposed by complex non-Markovian noise. By analyzing the control Hamiltonian for dynamically decoupled quantum sensing, we derived the necessary and sufficient conditions required to average out environmental effects, thereby recovering Heisenberg scaling with the QFI scaling as $t^2$. Notably, this result also holds despite initial correlations between the system and the environment, which may violate CP dynamics. These correlations give rise to quantum memory effects and, consequently, non-Markovian noise. Unlike Quantum Error Correction (QEC), which is predominantly effective for Markovian noise~\cite{zhou2018achieving}, DDT is applicable to both Markovian and non-Markovian regimes. This broader applicability underscores DDT’s potential as a powerful and efficient tool in real-world quantum sensing scenarios characterized by complex noise.

To illustrate our findings, we applied DDT to frequency estimation in the damped Jaynes–Cummings model, showing that it effectively preserves measurement accuracy in environments exhibiting pronounced quantum memory effects. This highlights DDT’s capability to mitigate environmental decoherence while preserving key quantum resources such as entanglement and superposition. Consequently, our results represent a step forward in optimizing quantum measurements, emphasizing the critical role of quantum control in advancing quantum metrology and sensing technologies.

Several important questions remain. Our focus here was primarily on achieving Heisenberg-limit scaling in time ($\sim t^2$). It would be valuable to determine how the approach presented here can be generalized to $N$ correlated systems and how the QFI behaves with respect to $N$. In particular, is it possible to retrieve the $\sim N^2$ QFI scaling for the $N$ correlated systems scenario using DDT? Furthermore, generalizing this approach to cases where the signal Hamiltonian $H_S(\omega)$ is a nonlinear function of $\omega$, as considered in this paper, warrants further investigation. These and related problems will likely guide the continued development of quantum sensing methods, especially for realistic implementations subject to environmental noise.

\section{Acknowledgments }

K.C.T. acknowledges support by the National Natural Science Fund of China (Grant No. G0512250610191).

\begin{appendices}
\appendix
\section{\MakeUppercase {Characterization of the control Hamiltonian for dynamically decoupled quantum sensing}} \label{App. A}

We prove a set of necessary and sufficient conditions for the existence of a dynamical decoupling process described by a set of control unitary operations $\{ U\left(kT\right) \}$ that satisfies Eq. \ref{Eq. 15}, and preserves the signal part of the Hamiltonian $H_S$ in the sense that the effective Hamiltonian $H^{\text{eff}}_S$ remains nontrivial after the dynamical decoupling process (Eq.~\ref{Eq. 15-2}). The following is Theorem~\ref{th::iff} in the main text.

\begin{theorem*} [Necessary and sufficient conditions for dynamical decoupled quantum sensing] 
    For any system-environment interaction $H_S \otimes \mathbb{1}_E + \mathbb{1}_S \otimes H_E + H_{SE}$, there exists a collection of unitary operators $\{ {U(kT)} \}_{k=0}^{n-1}$ that satisfies 
    \begin{equation*}
        H^{\text{eff}}_S = \frac{1}{n}\sum_{k=0}^{n-1} U(kT) \, H_S \, U(kT)^\dagger \not\propto \mathbb{1}_S
    \end{equation*} 
    and 
    \begin{align*}
       H_{SE}^{\text{eff}} &= \frac{1}{n}\sum_{k=0}^{n-1}\left( U_C(kT)\otimes \mathbb{1}_E \right) H_{SE} \left(U_C(kT)^\dagger \otimes \mathbb{1}_E\right) \\&= c \mathbb{1}_S \otimes J_E
    \end{align*}
    for some constant $c$ and Hermitian operator $J_E$ if and only if there exists mixed unitary channel $\Phi$ such that $\Phi(H_S)$ is nontrivial,
  \begin{equation*}
      \operatorname{Tr} \left\{\left[ \Phi(H_S) - \operatorname{Tr}(H_S) \frac{\mathbb{1}_S}{d} \right] \Phi \left[ \prescript{}{E}{\braket{\psi|H_{SE}|\psi}_E} \right] \right\}= 0
  \end{equation*}  
    and 
    \begin{equation*}
        \braket{i | \Phi \left[ \prescript{}{E}{\braket{\psi|H_{SE}|\psi}_E} \right]| i} = \mathrm{constant},
    \end{equation*}
    for every eigenvector $\ket{i}$ of $\Phi(H_S)$ and state vector $\ket{\psi}_E$.
\end{theorem*}

\begin{proof} 
    ($\Rightarrow$) Assume that $H^{\text{eff}}_S =\frac{1}{n} \sum_{k=0}^{n-1} U(kT) \, H_S \, U(kT)^\dagger \not\propto \mathbb{1}_S$ and $H_{SE}^{\text{eff}} = \frac{1}{n}\sum_{k=0}^{n-1} \left(U(kT)\otimes \mathbb{1}_E\right) \, H_{SE} \, \left(U(kT)^\dagger \otimes \mathbb{1}_E\right) = c \mathbb{1}_S \otimes J_E$ for some constant $c$ and Hermitian operator $J_E$.

    Since the operator $H^{\text{eff}}_S - \operatorname{Tr}(H^{\text{eff}}_S)\frac{\mathbb{1}_S}{d}$ has trace zero, and we assumed $H^{\text{eff}}_S \not\propto \mathbb{1}_S$, $H^{\text{eff}}_S - \operatorname{Tr}(H^{\text{eff}}_S)\frac{\mathbb{1}_S}{d}$ must be nontrivial and  does not have any component along $\mathbb{1}_S$. Observe that $ \prescript{}{E}{\braket{\psi|H_{E}^{\text{eff}}|\psi}_E}  \propto \mathbb{1}_S,$ which only has a component along $\mathbb{1}_S.$ This implies that 
    \begin{equation}
        \operatorname{Tr} \left\{\left[ H^{\text{eff}}_S - \operatorname{Tr}(H^{\text{eff}}_S) \frac{\mathbb{1}_S}{d} \right] \prescript{}{E}{\braket{\psi|H_{SE}^{\text{eff}}|\psi}_E} \right\} = 0.
    \end{equation}
    Not that the above expression is nontrivial only because $H^{\text{eff}}_S =\frac{1}{n} \sum_{k=0}^{n-1} U(kT) \, H_S \, U(kT)^\dagger \not\propto \mathbb{1}_S$.  

    Define the map $\Phi(\cdot) = \frac{1}{n}\sum_{k=0}^{n-1}U(kT) \,(\cdot)\, U^\dagger(kT)$. Since $\Phi$ is a convex combination of unitary maps, it is a mixed unitary channel by definition and also a CPTP map. As $\Phi$ is trace preserving, we have $\operatorname{Tr}(H^{\text{eff}}_S) =  \operatorname{Tr}(\Phi(H_S)) = \operatorname{Tr}(H_S)$. By further substituting the expressions $H^{\text{eff}}_S = \Phi(H_S)$ and $\prescript{}{E}{\braket{\psi|H_{SE}^{\text{eff}}|\psi}_E} =  \Phi\left[  \prescript{}{E}{\braket{\psi|H_{SE}|\psi}_E}  \right]$, we obtain the expression
    \begin{equation}
        \operatorname{Tr} \left\{\left[ \Phi(H_S) - \operatorname{Tr}(H_S) \frac{\mathbb{1}_S}{d} \right] \Phi \left[ \prescript{}{E}{\braket{\psi|H_{SE}|\psi}_E} \right] \right\}= 0.
    \end{equation}
    
    Let $\left \{ \ket{i} \right \}_{i=0}^{d-1}$ be the eigenbasis of $\Phi(H_S)$. Since $ \Phi \left[ \prescript{}{E}{\braket{\psi|H_{SE}|\psi}_E} \right] \propto \mathbb{1}_S$ for every $\ket{\psi_E}$, we must have $\braket{i | \Phi \left[ \prescript{}{E}{\braket{\psi|H_{SE}|\psi}_E} \right]| i} = \mathrm{constant}$ for every $i = 0,\ldots, d-1$. Therefore, there exists a mixed unitary map $\Phi$ that satisfies the require conditions such that $\Phi(H_S)$ is nontrivial.

    ($\Leftarrow$) Assume that there exists some mixed unitary map $\Phi$ such that $\operatorname{Tr} \left\{\left[ \Phi(H_S) - \operatorname{Tr}(H_S) \frac{\mathbb{1}_S}{d} \right] \Phi \left[ \prescript{}{E}{\braket{\psi|H_{SE}|\psi}_E} \right] \right\}= 0$, $\braket{i | \Phi \left[ \prescript{}{E}{\braket{\psi|H_{SE}|\psi}_E} \right]| i} = \mathrm{constant}$ for every eigenvector $\ket{i}$ of $\Phi(H_S)$ given any state vector $\ket{\psi}_E$ such that $\Phi(H_S)$ is nontrivial (not proportional to the identity).

    Since $\Phi(H_S)$ is nontrivial, the traceless operator $ \Phi(H_S) - \operatorname{Tr}(H_S) \frac{\mathbb{1}_S}{d} $ is also nontrivial. Writing the operator in its eigenbasis $\left \{ \ket{i} \right \}_{i=0}^{d-1}$, we obtain a diagonal representation 
    \begin{equation}
        \Phi(H_S) - \operatorname{Tr}(H_S) \frac{\mathbb{1}_S}{d}  = \begin{pmatrix} \lambda_0 & 0 &\cdots  & 0\\ 
    0 & \lambda_1 & \cdots & 0 \\ 
    \vdots & &\ddots & \vdots \\
    0 & 0 &\cdots& \lambda_{d-1} \end{pmatrix},
    \end{equation}
    where $\lambda_i$, $i=0,\ldots, d-1$ are the eigenvalues of  $ \Phi(H_S) - \operatorname{Tr}(H_S) \frac{\mathbb{1}_S}{d} $.
    
    Consider the operator $\Phi \left[ \prescript{}{E}{\braket{\psi|H_{SE}|\psi}_E} \right]$. Since $\braket{i | \Phi \left[ \prescript{}{E}{\braket{\psi|H_{SE}|\psi}_E} \right]| i} = \mathrm{constant}$ for every $i = 0, \ldots, d-1$, its matrix representation when written in the basis $\left \{ \ket{i} \right \}_{i=0}^{d-1}$ has the form 
    \begin{equation}
        \Phi \left[ \prescript{}{E}{\braket{\psi|H_{SE}|\psi}_E} \right] = \begin{pmatrix} c & * &\cdots  & *\\ 
    * & c & \cdots & * \\ 
    \vdots & &\ddots & \vdots \\
    * & * &\cdots& c \end{pmatrix},
    \end{equation}
    for some constant $c$. In this basis, the leading diagonals of the matrix $\Phi \left[ \prescript{}{E}{\braket{\psi|H_{SE}|\psi}_E} \right]$ is constant.

    Define the pinching channel $\Pi(\cdot) = \sum_{i=0}^{d-1} \ketbra{i} (\cdot) \ketbra{i}$. Pinching channels are defined as mixed, projective maps of the form $\Pi(\cdot) = \sum_i \Pi_i (\cdot) \Pi_i$ that satisfies $\sum_i \Pi_i = \mathbb{1}$ and $\Pi_i^2 = \Pi_i$. It is known that pinching channels are also mixed unitary channels, so for every pinching channel, there exists some distribution $p(x)$ over unitary matrices $U(x)$ such that 
\begin{equation}
    \Pi(\cdot) = \int dx \, p(x) U(x) (\cdot) U(x)^\dagger.
\end{equation}    
One may verify that the pinching channel satisfies $\Pi \circ \Phi (H_S) = \Phi (H_S)$ and $\Pi \circ \Phi \left[ \prescript{}{E}{\braket{\psi|H_{SE}|\psi}_E} \right] \propto \mathbb{1}_S $ given any $\ket{\psi}_E$. The second condition, in particular, implies that $(\Pi \circ \Phi)\otimes \mathbb{1}_E(H_{SE}) = c \mathbb{1}_S \otimes J_E$ for some Hermitian operator $J_E$. 

     Next, observe that since $\Phi$ and $\Pi$ are both mixed unitary channels, their composite $\Phi' = \Pi \circ \Phi$ is also a mixed unitary channel. This means that there exists some probability distribution density $p'(x)$  satisfying $\int dx \, p'(x)  = 1$ such that $\Phi'(\cdot) = \int p'\left(x\right)dx\, U(x) (\cdot) U(x)^\dagger.$ We can discretize the probability density by the following substitution 
\begin{equation}
    \int dx \, p'(x) \approx \sum_{j=0}^{m-1} p'(x_j) \Delta x \approx \sum_{j=0}^{m-1} \frac{m_j}{n},
\end{equation}     
where $\frac{m_j}{n} \approx p'(x_j) \Delta x$, and $n, m_j$ are nonnegative integers satisfying $\sum_{j=0}^{m-1} m_j = n$. This approximation is arbitrarily accurate for sufficiently large $n$.

     This allows us to write 
     \begin{equation}
         \Phi'(\cdot) =  \int p'\left(x\right)dx\, U(x) (\cdot) U(x)^\dagger = \sum_{j=0}^{m-1} \frac{m_j}{n} U(x_j) (\cdot) U(x_j)^\dagger .
     \end{equation}
Assume that $p'(x)$ is the uniform distribution over the range $[a,b]$ and the sampling points in this range, $x_k=a+k \Delta x$ correspond to equidistant values, where the duration of each small interval $\Delta x =T=(b-a)/n$,  and $k=0,1, \cdots, n-1$.  For any uniform
     distribution, the weights $m_k/n$ are equal for all bins and reduced to $1/n$. Therefore, one can rewrite the composite channel $\Phi' (\cdot)$ as 
\begin{equation}
         \Phi'(\cdot)  = \frac{1}{n}\sum_{k=0}^{n-1} U(kT) (\cdot) U(kT)^\dagger.
\end{equation}

Putting it all together, perform the substitution $n \Phi'(\cdot) \rightarrow \sum_{k=0}^{n-1} U(kT) (\cdot) U(kT)^\dagger$ to get  $H^{\mathrm{eff}}_S = \frac{1}{n}\sum_{k=0}^{n-1} U(kT) H_S U(kT)^\dagger \not\propto \mathbb{1}_S $ since $\Phi'(H_S) \not\propto \mathbb{1}$, and $H_{SE}^{\text{eff}} =\frac{1}{n} \sum_{k=0}^{n-1} \left(U(kT)\otimes \mathbb{1}_E\right) \, H_{SE} \, \left(U(kT)^\dagger \otimes \mathbb{1}_E\right) = c \mathbb{1}_S \otimes J_E$ for some constant $c$ and Hermitian operator $J_E$,  since $\Phi'\otimes \mathbb{1}_E (H_{SE}) \propto \mathbb{1}_S \otimes J_E$ for some Hermitian operator $J_E$. These are the required conditions and complete the proof.

     \end{proof}
     Theorem~\ref{th::iff} provides a characterization of dynamical decoupling in terms of the existence of mixed unitary channels that satisfies orthogonality conditions. Since the identity channel $\Phi(\cdot) = (\cdot)$ which leaves the input operator unchanged is a special case of a mixed unitary operator, the following corollary (Corollary~\ref{cor::if} in the main text) provides a simpler sufficient condition expressed in terms of the initial input Hamiltonian: 

\begin{corollary*} 
    There exists a collection of unitary operators $\{ {U(kT)} \}_{k=0}^{n-1}$ that satisfies 
    \begin{equation*}
        H^{\text{eff}}_S = \frac{1}{n}\sum_{k=0}^{n-1} U(kT) \, H_S \, U(kT)^\dagger \not\propto \mathbb{1}_S
    \end{equation*}
    and 
    \begin{align*}
       H_{SE}^{\text{eff}} &= \frac{1}{n}\sum_{k=0}^{n-1}\left( U_C(kT)\otimes \mathbb{1}_E \right) H_{SE} \left(U_C(kT)^\dagger \otimes \mathbb{1}_E\right) \\&= c \mathbb{1}_S \otimes J_E
    \end{align*}
    for some constant $c$ and Hermitian operator $J_E$, if $H_S$ is nontrivial and  
    \
    \begin{equation*}
        \operatorname{Tr} \left\{\left[ H_S - \operatorname{Tr}(H_S) \frac{\mathbb{1}_S}{d} \right]   \prescript{}{E}{\braket{\psi|H_{SE}|\psi}_E}  \right\}= 0
    \end{equation*}
    and 
    \begin{equation*}
        \braket{i | \prescript{}{E}{\braket{\psi|H_{SE}|\psi}_E} | i} = \mathrm{constant},
    \end{equation*}
    for every eigenvector $\ket{i}$ of $H_S$ and state vector $\ket{\psi}_E$.
\end{corollary*}

We demonstrate the usefulness of Corollary~\ref{cor::if} by discussing a simple example. Suppose the system and environment are composed of only two qubits. Let $H_S = \omega \sigma_z  $ and $H_{SE} = \sigma_x \otimes \sigma_x$ such that the total Hamiltonian is $H = H_S  + H_{SE}$. In quantum sensing, the goal is to preserve as much as possible the part of the Hamiltonian that encodes $\omega$ (the $H_S$ term), while eliminating as much as possible the information which does not encode $\omega$ (the $H_{SE}$ term). Let the eigenvectors of $H_S$ be denoted $\ket{i}$, where $i=0,1$. In this basis  
\begin{equation}
    H_S = \omega \begin{pmatrix}
    1 & 0 \\
    0 & -1
\end{pmatrix}
\end{equation}
is a diagonal matrix. On the other hand, we see that $\bra{i} \bra{\psi} H_{SE} \ket{i} \ket{\psi} = \braket{i|\sigma_x |i} \braket{\psi|\sigma_x |\psi} = 0$  for every $\ket{\psi}$. In matrix form, $H_{SE}$ is written as 
\begin{equation}
    H_{SE} = \begin{pmatrix} 0 & 1 \\ 1 & 0
\end{pmatrix} \otimes \sigma_x = \begin{pmatrix} 0 & \sigma_x \\ \sigma_x & 0 \end{pmatrix},
\end{equation}
so that the leading diagonal terms are the zero constant. As a result, we can completely eliminate the $H_{SE}$ term by projecting the first subsystem onto the eigenbasis of $H_S$ via the pinching channel $\Pi(\cdot) = \ketbra{0} (\cdot)  \ketbra{0} + \ketbra{1} (\cdot)  \ketbra{1}$, so that $\Pi \otimes \mathbb{1} (H_{SE}) = 0$. Without going into the exact implementation, which in general is not unique, we know that the pinching channel is a mixed unitary channel that can be mapped to a collection of unitary operators $\{ U(kT) \}_k$ such that $H_{SE}^{\mathrm{eff}} = 0$ (see proof of Theorem~\ref{th::iff}). For the purpose of quantum sensing eliminating the $H_{SE}$ term is not enough, as we still need to ensure that information about $\omega$ is not lost in the dynamical decoupling process. This is guaranteed by the condition 
\begin{widetext}
    \begin{equation}
    \operatorname{Tr} \left\{\left[ H_S - \operatorname{Tr}(H_S) \frac{\mathbb{1}_S}{d} \right]   \prescript{}{E}{\braket{\psi|H_{SE}|\psi}_E}  \right\} = \omega \braket{\psi|\sigma_x |\psi} \operatorname{Tr}( \sigma_z   \sigma_x  ) = 0,
\end{equation}
\end{widetext}
which says that the signal generated by $H_S$ is orthogonal to the noise generated by $H_{SE}$, so the signal can be preserved after dynamically decoupling away the $H_{SE}$ term. Indeed, we see that the pinching channel does not affect the signal, since $\Pi(H_S) = H_S$, so the encoding of $\omega$ is completely preserved.

\section{\MakeUppercase{Solving the Schrödinger Equation for a Two-Level System Embedded in a $k$-Mode Photonic Cavity}} \label{App. C}
Let's start with the Hamiltonian in the rotating wave approximation
\begin{equation}\label{HTLS}
    H_{tot}=\frac{\omega_0}{2}\sigma_{z}+\sum_{k} \omega_k a_k^{\dagger}a_k + \left(\sigma_{+}B +\sigma_{-}B^{\dagger}\right),
\end{equation}
where $B=\sum_ k g_k a_k$, with $a_k$ and $a_k^{\dagger}$ are photonic field annihilation and creation operators, respectively. $\sigma_z$ is the Pauli matrix and $\sigma_{+}$ and $\sigma_{-}$ are the raising and lowering operators for the atomic system, respectively.  $\omega_0$ and $\omega_k$ denote the atomic transition frequency and the $k$-boson mode frequency, respectively, while $g_k$ represents the coupling strength between the $k$-th field mode and the qubit. We consider the initial joint system-environment state as the single-excitation wave function, given by 
\begin{equation}
   \ket{\psi(0)}_{SE} = C_e(0) \ket{e 0} + C_g(0) \ket{g 0} + \sum_k C_k(0) \ket{g 0_k},
\end{equation}
where $C_e\left(0\right)=C_g\left(0\right)=1/\sqrt{2}$ representing the initial amplitudes of the excited and ground states, respectively. The time evolution of $ \ket{\psi(0)}_{SE}$ is given by 
\begin{equation}
   \ket{\psi(t)}_{SE} = C_e(t) \ket{e 0} + C_g(t) \ket{g 0} + \sum_k C_k(t) \ket{g 1_k},
\end{equation}
Since the evolution of the joint system-environment is unitary and governed by Hamiltonian \ref{HTLS}, the time-dependent amplitudes $C_e(t)$, $C_g(t)$ and $C_k(t)$ are determined by the Schrödinger equation and satisfy the following equations of motion
\begin{equation}\label{Eq. C4}
i \dot C_e(t)=\frac{\omega_0}{2}C_e(t)+\sum_k g_k C_k(t),
\end{equation}
\begin{equation}\label{Eq. C5}
i \dot C_g(t)=-\frac{\omega_0}{2}C_g(t),
\end{equation}
\begin{equation}\label{Eq. C6}
i \dot C_k(t)=\left(\omega_k-\frac{\omega_0}{2}\right)C_k(t)+g^{*}_k C_e(t).
\end{equation}
The amplitude $C_g(t)$ can be obtained directly from Eq. \ref{Eq. C5}, yielding $C_g(t) = C_g(0) e^{i \frac{\omega_0}{2} t}$ where $C_g(0)=1/\sqrt{2}$ is the initial amplitude. To solve Eqs. \ref{Eq. C4}, and \ref{Eq. C6} it is helpful to move to a rotating frame, where we define $C_e(t) = c_e(t) e^{-i \frac{\omega_0}{2} t}$ and $C_k(t) = c_k(t) e^{-i (\omega_k - \frac{\omega_0}{2}) t}$. This transformation removes the oscillations associated with the qubit frequency $\omega_0$ and the cavity mode frequency $\omega_k$, simplifying the analysis of the time-dependent amplitudes $c_g(t)$ and $c_k(t)$ in the rotating frame, and leads to
\begin{equation}\label{Eq. C7}
    i \dot c_e(t)=\sum_k g_k c_k(t) e^{i\left(\omega_0-\omega_k\right)t},
\end{equation}
\begin{equation}
    i \dot c_k(t)= g_k^{*} c_e(t) e^{-i\left(\omega_0-\omega_k\right)t}.
\end{equation}
Since there is initially no photon in the cavity, i.e. $c_k(0)=0$, then the equation for the amplitude $c_k(t)$ can be solved, yielding 
\begin{equation}
    c_k(t)= -i \int_0^{t} g_k^* c_e(\tau)e^{-i\left(\omega_0-\omega_k\right)t}.
\end{equation}
Substituting this solution into Eq. \ref{Eq. C7} yields the final integro-differential equation for $c_e(t)$
\begin{equation}
\dot c_e(t)=-\int_0^{t} f\left(t-\tau\right) c_e(\tau),
\end{equation}
where $f(t-\tau)$ is the kernel correlation function, which is directly related to the spectral density of the reservoir, $J(\omega)$, through a shifted Fourier transform $f(t-\tau) = \int J(\omega) e^{-(\omega - \omega_k)(t-\tau)} \, d\omega$.
Therefore, the exact solution for $c_e(t)$ depends on the specific choice of $J(\omega)$, which we assume has a Lorentzian form with detuning. For this particular choice, the correlation function $f(t-\tau)$ takes the form
\begin{equation}
  f(t-\tau) = \frac{1}{2} \gamma_0 \lambda e^{-(\lambda - i \Delta)(t - \tau)}.
\end{equation}
Using this correlation function along with the Laplace transform, one can obtain the exact solution for the amplitude $c_e(t)$, given by
\begin{equation}\label{Eq. C12}
    c_e(t)=c_e(0) e^{-i\left(\frac{\lambda-i\Delta}{2}\right)t}\left(\cosh{\frac{dt}{2}}+\frac{\lambda-i\Delta}{d}\sinh{\frac{dt}{2}}\right),
\end{equation}
where $c_e(0)=1/\sqrt{2}$ is the initial amplitude of $c_e(t)$, and $d=\sqrt{\left(\lambda-i \Delta\right)^2-2\gamma_0 \lambda}$. This amplitude of the excited state and its time derivative can be used to determine the time-dependent decay rate, $\gamma(t)=-2\Re\left[\dot c_e\left(t\right)/c_e\left(t\right)\right]$, and the time-dependent Lamb shift, $S(t)=-\Im \left[\dot c_e\left(t\right)/c_e\left(t\right)\right]$. $\gamma(t)$ describes the time-dependent probability decay of the system’s excited state and reveals how coherence is lost due to environmental interactions. In systems exhibiting non-Markovian behavior, $\gamma(t)$  can become negative over certain intervals, reflecting the “backflow” of information from the environment to the system. This backflow temporarily restores coherence, a distinctive feature of non-Markovian dynamics where memory effects play a significant role in the evolution of the quantum state. Returning to the Schrödinger picture, we can express the final state of the combined quit-environment system as follows
\begin{eqnarray}
    \begin{aligned}
         \ket{\psi\left(t\right)}_{SE}=& e^{-i\frac{\omega_0}{2} t}c_e(t)\ket{e 0}+\frac{e^{i\frac{\omega_0}{2}}}{\sqrt{2}}\ket{g 0} \\&
         +\sum_k e^{-i\left(\omega_k-\frac{\omega_0}{2}\right)} c_k(t)\ket{g 1_k},
    \end{aligned}
\end{eqnarray}
which represents the exact solution of the Schrödinger equation for a two-level system embedded in a $k$-mode photonic cavity.
\begin{widetext}
\section{\MakeUppercase{General Solution of Equation (\ref{Eq. Soce}) in the Main Text}} 
\label{App. D}
In this appendix, we provide a detailed derivation of the solution to the second-order differential equation (Eq. \ref{Eq. SecondDce}), which governs the amplitude of the excitation state in the presence of quantum control. We carefully outline each step of the derivation, beginning with the general form of the equation and progressing to the final solution that incorporates the effects of the applied quantum control. Indeed, the solution of Eq. \ref{Eq. SecondDce} takes the following general form
\begin{equation}\label{Eq. D1}
    c_e(t)=e^{-\left(\frac{\lambda-i\Delta}{2}\right)t}\left(A \cosh{\left[\frac{d t}{2}\right]}+B\sinh{\left[\frac{d t}{2}\right]}\right),
\end{equation}
where $d=\sqrt{\left(\lambda-i\Delta\right)^2-2\gamma_0\lambda}$, and the coefficients $A$ and $B$ are 

    \begin{equation}\label{Eq. D2}
    \small     A = e^{\left( {\frac{{\lambda  - i\Delta }}{2}} \right)\left( {n - 1} \right)T}\left( {\left( {\cosh \left[ {\frac{{d\left( {n - 1} \right)T}}{2}} \right] - \frac{{\lambda  - i\Delta }}{d}\sinh \left[ {\frac{{d\left( {n - 1} \right)T}}{2}} \right]} \right){c_e}\left( {\left( {n - 1} \right)T} \right) - \frac{2}{d}\sinh \left[ {\frac{{d\left( {n - 1} \right)T}}{2}} \right]{{\dot c}_e}\left( {\left( {n - 1} \right)T} \right)} \right),
    \end{equation}
     \begin{equation}\label{Eq. D3}
        B = e^{\left( {\frac{{\lambda  - i\Delta }}{2}} \right)\left( {n - 1} \right)T}\left( {\left( {\frac{{\lambda  - i\Delta }}{d}\cosh \left[ {\frac{{d\left( {n - 1} \right)T}}{2}} \right] - \sinh \left[ {\frac{{dTn}}{2}} \right]} \right){c_e}\left( {\left( {n - 1} \right)T} \right) + \frac{2}{d}\cosh \left[ {\frac{{d\left( {n - 1} \right)T}}{2}} \right]{{\dot c}_e}\left( {\left( {n - 1} \right)T} \right)} \right),
     \end{equation}
By inserting Eqs. \ref{Eq. D2} and \ref{Eq. D3} into Eq. \ref{Eq. D1}, one can obtain

    \begin{eqnarray*}
    \begin{aligned}
c_e\left( t \right) = & e^{ - \left( {\frac{{\lambda  - i\Delta }}{2}} \right)\left( {t - \left( {n - 1} \right)T} \right)}\left( {\cosh \left[ {\frac{{d\left( {t - \left( {n - 1} \right)T} \right)}}{2}} \right] + \frac{{\lambda  - i\Delta }}{d}\sinh \left[ {\frac{{d\left( {t - \left( {n - 1} \right)T} \right)}}{2}} \right]} \right){c_e}\left( {\left( {n - 1} \right)T} \right)\\&
+ \frac{2}{d}{e^{ - \left( {\frac{{\lambda  - i\Delta }}{2}} \right)\left( {t - \left( {n - 1} \right)T} \right)}}\sinh \left[ {\frac{{d\left( {t - \left( {n - 1} \right)T} \right)}}{2}} \right]{{\dot c}_e}\left( {\left( {n - 1} \right)T} \right).
\end{aligned}
    \end{eqnarray*}
This solution is valid only within the time interval $t\in\left[(n-1)T, nT\right]$, and depends on the boundary conditions $c_e((n-1)T_{+}) = c_e((n-1)T_{-})$ and $\dot{c}_e((n-1)T_{+}) = -\dot{c}_e((n-1)T_{-})$. By incorporating these conditions and using the linear decomposition, we can rewrite 
    \begin{equation}
      \left( {\begin{array}{*{20}{c}}
{{c_e}\left( t \right)}\\
{{{\dot c}_e}\left( t \right)}
\end{array}} \right) = {e^{ - \left( {\frac{{\lambda  - i\Delta }}{2}} \right)\left( {t - \left( {n - 1} \right)T} \right)}}\left( {\begin{array}{*{20}{c}}
{\alpha \left( {t,\left( {n - 1} \right)T} \right)}&{\delta \left( {t,\left( {n - 1} \right)T} \right)}\\
{\varepsilon \left( {t,\left( {n - 1} \right)T} \right)}&{\beta \left( {t,\left( {n - 1} \right)T} \right)}
\end{array}} \right)\left( {\begin{array}{*{20}{c}}
{{c_e}\left( {\left( {n - 1} \right)T_{-}} \right)}\\
{{{\dot c}_e}\left( {\left( {n - 1} \right)T_{-}} \right)}
\end{array}} \right),
    \end{equation}
    where the matrix elements are
    \begin{equation*}
        {\alpha \left( {t,\left( {n - 1} \right)T} \right) = \cosh \left[ {\frac{{d\left( {t - \left( {n - 1} \right)T} \right)}}{2}} \right] + \frac{{\lambda  - i\Delta }}{d}\sinh \left[ {\frac{{d\left( {t - \left( {n - 1} \right)T} \right)}}{2}} \right]},
    \end{equation*}
    \begin{equation*}
\beta \left( {t,\left( {n - 1} \right)T} \right) = \frac{{\lambda  - i\Delta }}{d}\sinh \left[ {\frac{{d\left( {t - \left( {n - 1} \right)T} \right)}}{2}} \right] - \cosh \left[ {\frac{{d\left( {t - \left( {n - 1} \right)T} \right)}}{2}} \right],
\end{equation*}
    \begin{equation*}
        {\delta \left( {t,\left( {n - 1} \right)T} \right) = -\frac{2}{d}\sinh \left[ {\frac{{d\left( {t - \left( {n - 1} \right)T} \right)}}{2}} \right]}, \quad  {\varepsilon \left( {t,\left( {n - 1} \right)T} \right) = \left( {\frac{d}{2} - \frac{{{{\left( {\lambda  - i\Delta } \right)}^2}}}{{2d}}} \right)\sinh \left[ {\frac{{d\left( {t - \left( {n - 1} \right)T} \right)}}{2}} \right]},
    \end{equation*}
    and 
    \begin{equation*}
        \left( {\begin{array}{*{20}{c}}
{{c_e}\left( {\left( {n - 1} \right)T} \right)}\\
{{{\dot c}_e}\left( {\left( {n - 1} \right)T} \right)}
\end{array}} \right) = {e^{ - \left( {\frac{{\lambda  - i\Delta }}{2}} \right)T}}{\left. {\left( {\begin{array}{*{20}{c}}
{\alpha \left( {t,\left( {n - 1} \right)T} \right)}&{\delta \left( {t,\left( {n - 1} \right)T} \right)}\\
{\varepsilon \left( {t,\left( {n - 1} \right)T} \right)}&{\beta \left( {t,\left( {n - 1} \right)T} \right)}
\end{array}} \right)} \right|_{t = nT}}\left( {\begin{array}{*{20}{c}}
{{c_e}\left( {\left( {n - 2} \right)T} \right)}\\
{{{\dot c}_e}\left( {\left( {n - 2} \right)T} \right)}
\end{array}} \right).
    \end{equation*}
    After applying $n-1$-sequential controls and using the recurrence relation, one can easily find
    \begin{equation}\label{Eq. D5}
        \left( {\begin{array}{*{20}{c}}
{{c_e}\left( t \right)}\\
{{{\dot c}_e}\left( t \right)}
\end{array}} \right) = {e^{ - \left( {\frac{{\lambda  - i\Delta }}{2}} \right)t}}\left( {\begin{array}{*{20}{c}}
{\alpha \left( {t,\left( {n - 1} \right)T} \right)}&{\delta \left( {t,\left( {n - 1} \right)T} \right)}\\
{\varepsilon \left( {t,\left( {n - 1} \right)T} \right)}&{\beta \left( {t,\left( {n - 1} \right)T} \right)}
\end{array}} \right){\mathcal{M}^{n - 1}}\left( {\begin{array}{*{20}{c}}
{{c_e}\left( 0 \right)}\\
0
\end{array}} \right),
    \end{equation}
        remind that $\dot c_e(0)=0$, and 
\begin{equation*}
\mathcal{M} = {\left. {\left( {\begin{array}{*{20}{c}}
{\alpha \left( {t,\left( {n - 1} \right)T} \right)}&{\delta \left( {t,\left( {n - 1} \right)T} \right)}\\
{\varepsilon \left( {t,\left( {n - 1} \right)T} \right)}&{\beta \left( {t,\left( {n - 1} \right)T} \right)}
\end{array}} \right)} \right|_{t = nT}} = \left( {\begin{array}{*{20}{c}}
{\cosh \left[ {\frac{{dT}}{2}} \right] + \frac{{\lambda  - i\Delta }}{d}\sinh \left[ {\frac{{dT}}{2}} \right]}&{ - \frac{2}{d}\sinh \left[ {\frac{{dT}}{2}} \right]}\\
{\left( {\frac{d}{2} - \frac{{{{\left( {\lambda  - i\Delta } \right)}^2}}}{{2d}}} \right)\sinh \left[ {\frac{{dT}}{2}} \right]}&{\frac{{\lambda  - i\Delta }}{d}\sinh \left[ {\frac{{dT}}{2}} \right] - \cosh \left[ {\frac{{dT}}{2}} \right]}
\end{array}} \right).
\end{equation*}
To evaluate $M^{n-1}$, $M$ must be diagonalizable. This means there must exist a matrix $P$ and a diagonal matrix $D$ such that $P^{-1} M P = D$. Then, $M^{n-1}$ can be expressed as $P D^{n-1} P^{-1}$. The elements of matrices $P = \left( {\begin{array}{*{20}{c}}
{{P_{11}}}&{{P_{12}}}\\
{{P_{21}}}&{{P_{22}}}
\end{array}} \right)$ and $D=\operatorname{Diag}(D_{11},D_{22})$ are given, respectively, as 
${{P_{11}} = \frac{{2\csch\left[ {\frac{{dT}}{2}} \right]\left( { - d\cosh \left[ {\frac{{dT}}{2}} \right] + \sqrt {{d^2} - {{\left( {\Delta  + i\lambda } \right)}^2}\sinh {{\left[ {\frac{{dT}}{2}} \right]}^2}} } \right)}}{{{d^2} + {{\left( {\Delta  + i\lambda } \right)}^2}}}}$, $P_{21}=P_{22}=1$ and ${P_{12}} =  - \frac{{2\csch\left[ {\frac{{dT}}{2}} \right]\left( {d\cosh \left[ {\frac{{dT}}{2}} \right] + \sqrt {{d^2} - {{\left( {\Delta  + i\lambda } \right)}^2}\sinh {{\left[ {\frac{{dT}}{2}} \right]}^2}} } \right)}}{{{d^2} + {{\left( {\Delta  + i\lambda } \right)}^2}}}$, ${D_{11}} = \frac{{\left( {\lambda-i\Delta } \right)\sinh \left[ {\frac{{dT}}{2}} \right] - \sqrt {{d^2} - {{\left( {\Delta  + i\lambda } \right)}^2}\sinh {{\left[ {\frac{{dT}}{2}} \right]}^2}} }}{d}$ and  ${D_{22}} = \frac{{\left( {\lambda  - i\Delta } \right)\sinh \left[ {\frac{{dT}}{2}} \right] + \sqrt {{d^2} - {{\left( {\Delta  + i\lambda } \right)}^2}\sinh {{\left[ {\frac{{dT}}{2}} \right]}^2}} }}{d}$. Hence, we have 
\begin{equation}\label{Eq. D6}
\mathcal{M}^{n - 1} = \left( {\begin{array}{*{20}{c}}
{{P_{11}}}&{{P_{12}}}\\
{{P_{21}}}&{{P_{22}}}
\end{array}} \right)\left( {\begin{array}{*{20}{c}}
{{{\left( {{D_{11}}} \right)}^{n - 1}}}&0\\
o&{{{\left( {{D_{22}}} \right)}^{n - 1}}}
\end{array}} \right){\left( {\begin{array}{*{20}{c}}
{{P_{11}}}&{{P_{12}}}\\
{{P_{21}}}&{{P_{22}}}
\end{array}} \right)^{ - 1}}.
\end{equation}
By substituting Eq. \ref{Eq. D6} into Eq. \ref{Eq. D5}, one can obtain
    \begin{equation*}
  \small  \left( {\begin{array}{*{20}{c}}
{{c_e}\left( t \right)}\\
{{{\dot c}_e}\left( t \right)}
\end{array}} \right) = {e^{ - \left( {\frac{{\lambda  - i\Delta }}{2}} \right)t}}\left( {\begin{array}{*{20}{c}}
{\alpha \left( {t,\left( {n - 1} \right)T} \right)}&{\delta \left( {t,\left( {n - 1} \right)T} \right)}\\
{\varepsilon \left( {t,\left( {n - 1} \right)T} \right)}&{\beta \left( {t,\left( {n - 1} \right)T} \right)}
\end{array}} \right)\left( {\begin{array}{*{20}{c}}
{{P_{11}}}&{{P_{12}}}\\
{{P_{21}}}&{{P_{22}}}
\end{array}} \right)\left( {\begin{array}{*{20}{c}}
{{{\left( {{D_{11}}} \right)}^{n - 1}}}&0\\
0&{{{\left( {{D_{22}}} \right)}^{n - 1}}}
\end{array}} \right){\left( {\begin{array}{*{20}{c}}
{{P_{11}}}&{{P_{12}}}\\
{{P_{21}}}&{{P_{22}}}
\end{array}} \right)^{ - 1}}\left( {\begin{array}{*{20}{c}}
{{c_e}\left( 0 \right)}\\
0
\end{array}} \right).
\end{equation*}
Therefore, the amplitude of the excitation state when $n$-sequential controls are applied is obtained as
\begin{equation}
    c_e\left( t \right) = {c_e}\left( 0 \right){e^{ - \frac{1}{2}\left( {\lambda  - i\Delta } \right)t}}\left( {{A_n}\cosh \left[ {\frac{1}{2}d\left( {t - nT} \right)} \right] + {B_n}\sinh \left[ {\frac{1}{2}d\left( {t - nT} \right)} \right]} \right), 
\end{equation}
where the coefficients $A_n$ and $B_n$ are given by
\begin{equation*}
    {A_n} = \frac{1}{{{K}}}\left( {{K} - d\cosh \left[ {\frac{{d T}}{2}} \right]} \right){\eta _ {-} }^n + \left( {{K} + d\cosh \left[ {\frac{{d T}}{2}} \right]} \right){\eta _ {+} }^n,
\end{equation*}
\begin{equation*}
    {B_n} = \frac{{d\left( {\eta _ + ^n + \eta _ - ^n} \right)\left( {\lambda  - {\rm{i}}\Delta } \right)\cosh \left[ {\frac{{dT}}{2}} \right] + \left( {\eta _ + ^n - \eta _ - ^n} \right)\left( {K\left( {\lambda  - {\rm{i}}\Delta } \right) - \left( {{d^2} + {{\left( {\Delta  + {\rm{i}}\lambda } \right)}^2}} \right)\sinh \left[ {\frac{{dT}}{2}} \right]} \right)}}{{2dK}},
\end{equation*}
with ${\eta _ \pm } = \frac{{\left( {\lambda  - {\rm{i}}\Delta } \right)\sinh \left[ {\frac{{dT}}{2}} \right] \pm {K}}}{d}$ and $K = \sqrt {{d^2} - {{\left( {\Delta  + {\rm{i}}\lambda } \right)}^2}\sinh {{\left[ {\frac{{dT}}{2}} \right]}^2}}$. Note that when $n= 0$, it can be easily shown that $A_0 = 1$ and $B_0 = \frac{\lambda - i\Delta}{d}$, which leads to the result in Eq. \ref{Eq. C12} corresponding to the amplitude $c_e(t)$ in the absence of control. 
\end{widetext}

\end{appendices}

\end{document}